\documentclass[reprint, superscriptaddress, amsmath,amssymb, aps, prb, floatfix]{revtex4-2}

\usepackage{comment}
\usepackage{dsfont}
\usepackage{graphicx}
\usepackage{dcolumn}
\usepackage{bm}
\usepackage{braket}
\usepackage{hyperref}
\usepackage{color}

\begin{document}

\preprint{APS/123-QED}

\title{Two-site Kitaev sweet spots evolving into topological islands}

\author{Rodrigo A. Dourado}
\email{dourado.rodrigo.a@gmail.com}
\affiliation{%
 \textit{Instituto de F\'isica de S\~ao Carlos, Universidade de S\~ao Paulo, 13560-970 S\~ao Carlos, S\~ao Paulo, Brazil}}

\author{J. Carlos Egues}
\affiliation{%
	\textit{Instituto de F\'isica de S\~ao Carlos, Universidade de S\~ao Paulo, 13560-970 S\~ao Carlos, S\~ao Paulo, Brazil}}

\author{Poliana H. Penteado}%
\email{polianahp@gmail.com}
\affiliation{%
 \textit{Instituto de F\'isica de S\~ao Carlos, Universidade de S\~ao Paulo, 13560-970 S\~ao Carlos, S\~ao Paulo, Brazil}}

\date{\today}

\begin{abstract}
Artificial Kitaev chains based on arrays of quantum dots are promising platforms for realizing Majorana Bound States (MBSs). In a two-site Kitaev chain, it is possible to find these non-Abelian zero-energy excitations at certain points in parameter space (sweet spots). These states, commonly referred to as Poor man's Majorana bound states (PMMs), are challenging to find and stabilize experimentally. In this work, we investigate the evolution of the sweet spots as we increase the number of sites of the Kitaev chain. To this end, we use the Bogoliubov-de Gennes representation to study the excitations of the system, and the scattering matrix and Green functions formalisms to calculate the zero-bias conductance. Our results show that the sweet spots evolve into a region that grows bigger and becomes gradually more protected as the number of sites $N$ increases. Due to the protection of the MBSs, we refer to this region as a topological island. We obtain similar results by considering a realistic spinful model with finite magnetic fields in a chain of normal-superconducting quantum dots. For long chains, $N \geq 20$, we show the emergence of strictly zero-energy plateaus robust against disorder. Finally, we demonstrate that the topological islands can be observed by performing conductance measurements via a quantum dot side-coupled to the Kitaev chain. Our work shows that the fine-tuning required to create and detect PMMs in a 2-site Kitaev chain is significantly relaxed as the length of the chain increases and details how PMMs evolve into MBSs. Our results are consistent with experimental reports for 2 and 3-site chains.
  
\end{abstract}

\maketitle

\section{Introduction}

The implementation of fault-tolerant quantum computation \cite{kitaev1, nayak2008non} has motivated an intense search for topological superconductors in semiconducting-superconducting hybrid structures over the past decades~\cite{mourik2012, leijnse2012introduction, aliceareview, beenakker2013search, nadj2014observation, chen2019ubiquitous, flensberg2021engineered, prada2020andreev,  van2023electrostatic}, including protocols for the detection of Majorana bound states (MBSs)~\cite{dourado2024nonlocality, nayakExperiment}. These zero-energy excitations of topological superconductors are predicted to present remarkable robustness against noise and obey non-Abelian braiding statistics~\cite{kitaev2, ivanov}.  Recently, promising experimental results consistent with the realization of artificial 2 and 3-site Kitaev chains in short arrays of quantum dots were reported \cite{dvir2023, bordin2024signatures, bordin2024crossed}, representing a significant step towards the demonstration of MBSs. 

The Kitaev chain was originally conceived as a toy model for spinless p-wave superconductors \cite{kitaev2, lieb1961two}. It has been predicted, though, that by coupling quantum dots (QDs) via short superconductors, it is possible to emulate Kitaev chains in realistic setups ~\cite{sau2012realizing, fulga2013adaptive}.
In recent developments of these proposals, Andreev Bound States (ABSs) localized in the superconducting dots couple normal dots via elastic cotunneling (ECT) and crossed Andreev reflection (CAR) \cite{liu2022tunable, tsintzis2022creating}. Once the superconductor is projected out, and provided that an applied magnetic field is strong enough to polarize the normal quantum dots, the effective Hamiltonian has the same form as the Kitaev model, with the ECT and CAR coefficients mapping onto the hopping and p-wave pairing amplitudes, respectively. The ratio between ECT and CAR can be tuned via a gate voltage applied to the superconducting dot, which in turn controls the energy level of the ABSs ~\cite{liu2022tunable, wang2022singlet, bordin2022controlled}. 

An artificial 2-site Kitaev chain can be experimentally realized with a pair of normal quantum dots coupled via an ABS~\cite{tsintzis2022creating, dvir2023}. In this system, one can find localized zero-energy states that share most of the MBSs' properties, including non-Abelian statistics and quantized conductance~\cite{leijnsepoorman}. However, these zero-energy excitations only exist at ``sweet spots" in parameter space, i.e., any fluctuation in the parameters can cause an energy splitting and/or their wave functions to overlap. The lack of robustness led these states to be commonly referred to as ``Poor man's Majoranas" (PMMs) \cite{leijnsepoorman}. The topological protection is achieved by increasing the number of quantum dots. This can already be observed in a 3-site Kitaev chain, where zero-bias conductance peaks have been shown to present robustness against parameter variations \cite{bordin2024signatures}. Therefore, it is important to characterize the crossover between short and long Kitaev chains to assess where the topological protection starts to play a role and at which point we can expect to observe MBSs in experiments. 

To address these questions, here we investigate the evolution of 2-site sweet spots as we increase the number of sites in the Kitaev model \cite{kitaev2}. We observe that in the short chain limit ($N < 10$) the sweet spot grows into a region stable against variations in all gate voltages. This result, Fig.~\ref{Fig1}, indicates that further extending the chain in current experiments \cite{dvir2023, ten2024two, bordin2024signatures, Haaf_arXiv2024} to $N = 4, 5$ might significantly ease the process of finding and stabilizing PMMs. We also include simulations of a microscopic model (normal QDs coupled via superconducting dots) at finite magnetic fields \cite{tsintzis2022creating} to corroborate this claim, Fig.~\ref{Fig2}. We increase the number of sites until we find regions in parameter space where the excitations' energy remains exactly zero even in the presence of disorder, Fig.~\ref{Fig4}. At this point, we observe a well-delimited region in which the MBSs are protected against local perturbations. We thus refer to such regions as topological islands.


As one of our main goals is to describe the evolution of discrete sweet spots into topological regions, it is essential to define the criteria we use to characterize such sweet spots. First, the ground state must be degenerate; this ensures that PMMs or MBSs are solutions of the Hamiltonian, which leads to self-adjoint Bogoliubov operators, and second, the wave functions of the zero-energy excitations must be well-localized at the edges of the system. To quantify the degree of localization, we use the Majorana Polarization (MP) \cite{sedlmayr2015visualizing, tsintzis2022creating}. A small value of MP indicates that the PMMs/MBSs are not well-localized and their wave functions overlap. Based on these two criteria, we investigate how the 2-site Kitaev sweet spot evolves in different regimes.

In the following, we briefly describe our main results.

\subsection{Short Kitaev chains}

We start by analyzing the 2-site Kitaev chain and incrementally adding sites to it. In a Kitaev chain with $N$ sites, there is a family of $N$ zero-energy solutions depicted by lines in parameter space \cite{grifoniSpectrum}. Similarly, there are $N-1$ lines along which the MP is maximized. All of these lines converge at the 2-site sweet spot. Hence, as we add more sites (lines) a sweet spot expands into a region in parameter space, see Fig.~\ref{Fig1}, where the energy and the MP assume minimum and maximum values, respectively. The area of this region scales with the number of sites. As we show in Fig.~\ref{Fig1}(d), this significantly improves the protection of the zero modes against variations in the parameters, which signals the beginning of a crossover from PMMs to MBSs. 

We also include simulations using a realistic model to describe an array of semiconducting-superconducting spinful QDs at finite magnetic fields \cite{tsintzis2022creating}. This model is predicted to emulate artificial Kitaev chains \cite{liu2022tunable, tsintzis2022creating} and reproduces signatures of PMMs observed in recent experiments \cite{dvir2023, ten2024two, bordin2024signatures}. The results for the spinful model show qualitatively similar features to the Kitaev chain, see Fig.~\ref{Fig2}. For instance, there are additional lines both in the energy and the MP plots when we increase the number of QDs in the system. The main finding is that the convergence of the zero-energy and high MP lines expands the 2-site sweet spot into a region. For both the Kitaev chain and the spinful model, for $N = 9$, we can already observe the emergence of a well-delimited region with zero-energy lines resembling the boundaries of a topological phase.

\subsection{Long Kitaev chains}

\begin{figure}
	\centering	\includegraphics[width=0.475\textwidth]{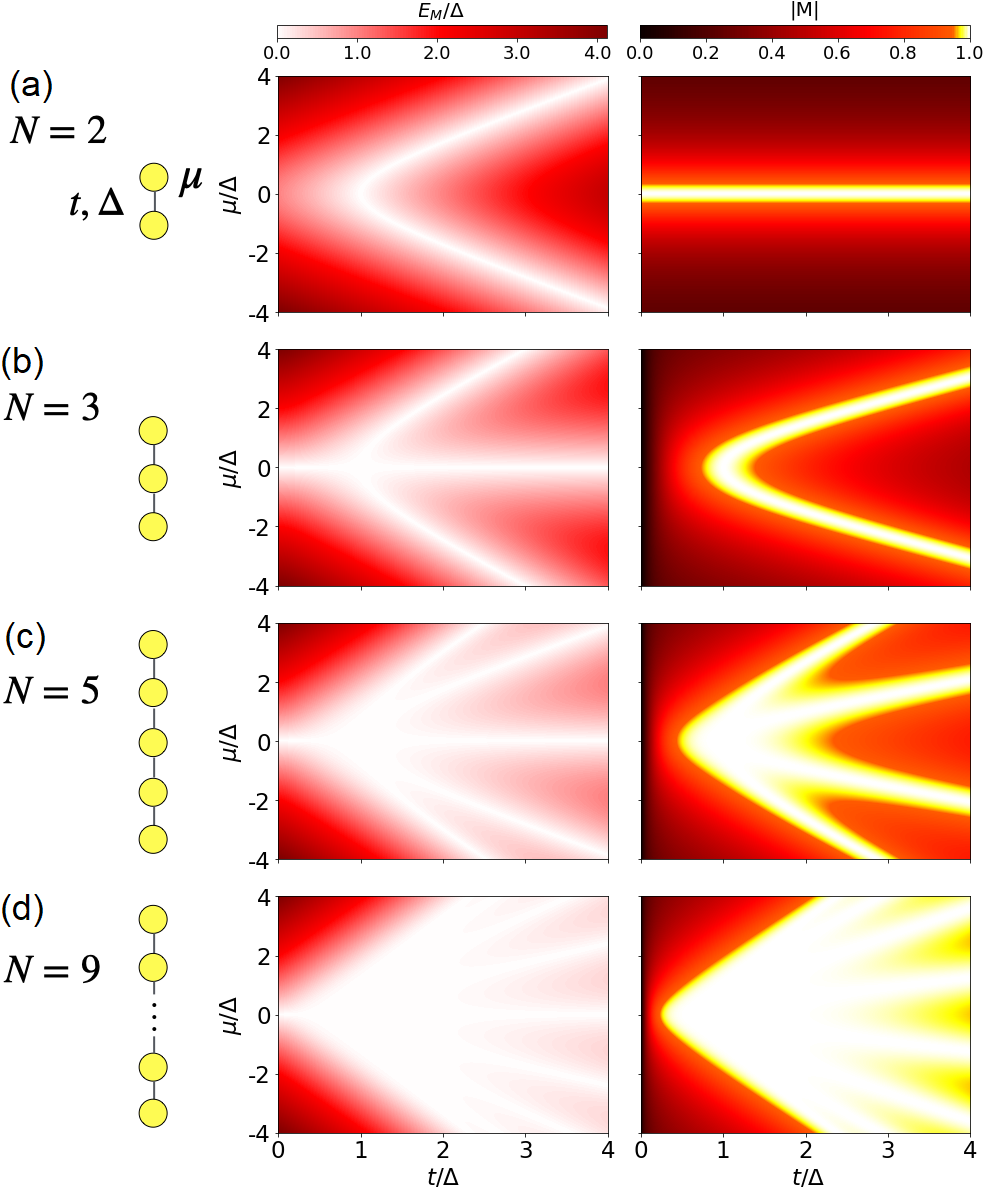}
	\caption{Lowest energy excitation of the Bogoliubov-de Gennes spectrum (left panel) and Majorana Polarization (right panel), as defined in Eq. (\ref{MP definition}), as functions of $\mu/\Delta$ and $t\Delta$ for an increasing number of sites. (a)-(d) $N=2, 3, 5, 9$, respectively. As $N$ increases, the convergence of zero-energy lines in the sweet spot causes the forming of a region where $E_M \to 0$. In addition, there are $(N-1)$ lines where $|M_1| = |M_N| = |M| = 1$ in parameter space crossing the sweet spot. The sweet spot grows into a region where $E_M \to$ and $|M| \to 1$ as $N$ increases.}
	\label{Fig1}
\end{figure}

In the limit of long chains, $N \geq 20$, we observe features consistent with topologically protected MBSs. We identify a region in parameter space around the 2-site sweet spot where the energy is strictly zero, as shown in Fig.~\ref{Fig4}, and verify that this region overlaps with well-localized MBSs, leading to maximized MP values. This overlap is already noticeable in the short chain limit for $N = 9$, see Figs.~\ref{Fig1}(d) and \ref{Fig2}(d). We test the robustness of our results by including an Anderson-type disorder in all parameters. We find that as the length of the chain increases, the zero-energy regions become essentially unaltered by disorder effects of the order of the gap, see Fig.~\ref{Fig4}. This robustness of the MBSs allows us to identify this region as a topological island.

\subsection{Conductance measurements}

We propose to observe the topological islands via conductance measurements. Our setup consists of a spin-polarized QD connected to metallic leads and side-coupled to a Kitaev chain.~\cite{barangerDot, vernekLeakage, David2015, weymann2017transport, souto2023probing}, Fig.~\ref{Fig5}(a). In this setup, the MBS leaks into the probe QD causing distinctive transport features such as an $e^2/2h$ quantized zero-bias conductance plateau~\cite{vernekLeakage}. We reproduce this result within the topological islands, Fig.~\ref{Fig5}(c). In addition, we can control the coupling of the MBS to the lead by tuning the dot level $\varepsilon_d$. By setting $\varepsilon_d \gg \Delta$, with $\Delta$ the superconducting pairing potential, the conductance becomes highly sensitive to the hybridization energy. At low temperatures ($T \to 0$), this allows for precise tracking of the zero-energy plateaus, see Fig.~\ref{Fig5}(b). At finite temperatures, we can still detect the topological island by choosing larger values of $\varepsilon_d$, albeit with limited resolution, see Fig.~\ref{Fig6}.

The paper is organized as follows. In Sec. \ref{modelHamiltonian}, we present the models used to describe our systems: the Kitaev chain and the Kitaev-QD setup. In Sec. \ref{degeneracy and MP}, we associate zero-energy solutions with self-Adjoint operators and define the Majorana Polarization. In Secs. \ref{scaling short chains} and \ref{spinfulmodel Section}, we consider short chains and calculate the energy and the MP for an increasing number of sites. In Sec \ref{long chains}, we study long chains ($N > 20$) and characterize the topological islands. In Sec. \ref{conductance section}, we calculate the conductance in the QD-Kitaev setup and propose a method of probing the zero-energy plateaus. We present our conclusions in Sec. \ref{conclusions}.

\section{Model Hamiltonians} \label{modelHamiltonian}

In the following, we describe the Hamiltonians we use in this work.

\subsection{Kitaev Model}

The Kitaev model describes 1D spinless p-wave superconductors \cite{kitaev2}. The Hamiltonian representing the system is given by 

\begin{equation} \label{KitaevChain}
    H_\mathrm{K} = - \sum_{j = 1}^N \mu c^\dagger_j c_j + \sum_{j = 1}^{N-1} \left(\Delta  c_j^\dagger c_{j+1}^\dagger - t c^\dagger_{j+1} c_{j} + H.c. \right),
\end{equation}
where $c_j$ is the annihilation of an electron on site $j$, $\mu$ is the chemical potential with respect to the grounded superconductor, $t$ is the hopping, and $\Delta$ is the p-wave pairing amplitude. As we shall see in Sec.~\ref{long chains}, we also consider possible fluctuations in the tuning of the parameters of the above Hamiltonian. We model these as an Anderson-type disorder. In this case, for instance, $\mu \to \mu_j = \mu - \delta \mu_j$, where $\delta \mu_j$ is assigned a random value in the interval $[-W, W]$, where $W$ is the disorder strength.

\subsection{Spinful model}

We can emulate a Kitaev chain by considering an array of normal QDs coupled by superconducting QDs \cite{sau2012realizing, tsintzis2022creating}. The Hamiltonian that describes this system reads 

\begin{equation} \label{Hspinful}
    \begin{split}
        H_{\text{QDs}} &= \sum_{j, \sigma} \left(\varepsilon_j + s_{\sigma} V_{z, j} \right) c_{j, \sigma}^\dagger c_{j, \sigma} + \sum_{j}\left[\delta_j c_{j, \uparrow}^\dagger c_{j, \downarrow}^\dagger + H.c. \right]\\
        &+\sum_{j, \sigma}\left[ t c_{j+1, \sigma}^\dagger c_{j, \sigma} + s_{\sigma} t^{so} c_{j+1, \sigma}^\dagger c_{j, \bar{\sigma}}+ H.c. \right],
    \end{split}
\end{equation}
where $\varepsilon_j$ is the on-site energy of the QDs that can be tuned via external gates, $V_{z, j}$ is the Zeeman energy associated with an externally applied magnetic field, with $s_\uparrow(s_\downarrow) = \pm 1$, $\delta$ represents an s-wave pairing amplitude, $t$ is the spin-conserving hopping, and $t^{so}$ is the spin-flip tunneling associated to a spin-orbit field along the y-direction and perpendicular to the applied magnetic field along the z-axis.~\cite{stepanenko2012singlet, tsintzis2022creating}. We consider $N$ normal QDs and $N-1$ superconducting QDs, such that $\delta_j = \delta$ when $j$ is even and $\delta_j = 0$ when $j$ is odd.

\subsection{Quantum dot coupled to the Kitaev chain}

We propose conductance measurements in the setup shown in Fig.~\ref{Fig5}(a), where we side couple a QD to the Kitaev chain \cite{vernekLeakage, barangerDot, David2015}. We also attach two metallic leads to the QD and measure the zero-bias conductance through the QD. The full Hamiltonian for this case is 

\begin{equation} \label{FullHdotKitaev}
    H_{\text{QD-Kitaev}} = \varepsilon_d c_d^\dagger c_d + t_0 \left(c_d^\dagger c_1 + H.c. \right) + H_K,
\end{equation}
where $c_d$ destroys an electron on the probe-QD, $\varepsilon_d$ is the on-site energy, and $t_0$ is the hopping parameter. 

To obtain analytic expressions, we also consider a low-energy description of the Kitaev chain~\cite{demler, PLeeMajoranaInducedAndreev}. The effective Hamiltonian for the QD-Kitaev system is then given by

\begin{equation} \label{dotKitaevH}
    H_{eff} =  \varepsilon_d c^\dagger_d c_d + t_0 \left(c_d^\dagger - c_d \right)  \gamma_1 + i E_M \gamma_1 \gamma_2,
\end{equation}
where $\gamma_i = \gamma_i^\dagger$, $\{\gamma_i, \gamma_j \} = 2\delta_{ij}$, are the Majorana operators and $E_M$ is the hybridization energy between them.  The tunneling Hamiltonian between the probe-QD and the metallic leads reads
\begin{equation} \label{leadsDot}
    H_T =  \sum_{k, j = 1, 2} t_{k,j} ( c^\dagger_d d_{k, j} + H.c.),
\end{equation}
where $d_{k, j}$ destroys an electron of momentum $k$ on lead $j$ and $t_{k, j}$ is the hopping amplitude.

\section{Degeneracy of the ground state and Majorana Polarization} \label{degeneracy and MP}

In this section, we detail the two quality factors that have been used in the literature to describe sweet spots, i.e., the degeneracy of the ground state and the Majorana Polarization~\cite{tsintzis2022creating, tsintzis2024majorana}. We show that when the energy of the first excitation is exactly zero (degeneracy point), the associated Bogoliubov operator naturally becomes a Majorana operator. In addition, we quantify the separation between the MBSs/PMMs using the MP, discussing its implication on the stability of the zero modes with respect to fluctuations in the parameters.

\subsection{Zero energy excitations}

In the following, we focus on the energy of the first excited state in the Bogoliubov-de Gennes (BdG) spectrum. This energy, $E_M$, can be interpreted as resulting from the hybridization between the two end Majorana modes (MBSs or PMMs), see Eq.~(\ref{dotKitaevH}). When $E_M = 0$, the creation of a fermionic excitation described by the operator $f^\dagger = (\gamma_1 + i \gamma_2)$ has no energy cost, meaning that $\ket{0}$ and $f^\dagger \ket{0}$ are degenerate ground states. The analytical diagonalization in Ref.~\cite{grifoniSpectrum} shows that for an $N$-site Kitaev chain, there is a family of $N$ solutions in parameter space for which the ground state is degenerate. The general analytical expression for these solutions is given by~\cite{grifoniSpectrum, grifoniConductance}

\begin{equation} \label{majoranaLines}
\mu^{(n)} = 2 \sqrt{t^2 - \Delta^2}\cos \left(\frac{n \pi}{N+1} \right), \, \, \, n = 1, 2, ..., N,
\end{equation}
where the parameters are considered homogeneous, see Eq.~(\ref{KitaevChain}). For the 2-site Kitaev chain, we have two lines describing zero-energy solutions given by $\mu^{(1, 2)} = \pm \sqrt{t^2 - \Delta^2}$, see Fig.~\ref{Fig1}(a). For each additional site, there is a corresponding line in parameter space as shown in the first column of Fig.~\ref{Fig1}. The hybridization energy increases as one moves away from the lines, reaching its maximum at the middle point between the two closest lines.

To identify ``sweet spots", we require ground state degeneracy, which implies $E_M = 0$, meaning that the parameters must be chosen to lie along one of the lines described by Eq.~(\ref{majoranaLines}). However, we find that by looking at the algebra of the associated Bogoliubov operators, rather than the value of $E_M$ itself, there is a clearer distinction between zero-energy and finite-energy states, which enhances our ability to characterize sweet spots.

To show this distinction formally, we next express the Bogoliubov operators in terms of the eigenfunctions and corresponding energies of the Bogoliubov Hamiltonian. We start by diagonalizing the Bogoliubov equation, $\mathcal{H}_{BdG}\phi_n = E_n \phi_n$, and obtaining the energies $E_n$ and eigenfunctions $\phi_n^\dagger = \begin{pmatrix}u_n & v_n \end{pmatrix}$, where $u_n = \begin{pmatrix}u_{1, n} & u_{2, n} & ... & u_{N, n} \end{pmatrix}$, with $u_{j, n} = u_j(E_n)$; a similar expression holds for $v_n$. Using this set of eigenfunctions, we define the unitary transformation $U = \begin{pmatrix} \phi_1 & ... & \phi_n & \phi_{n+1} & ... & \phi_{2N}  \end{pmatrix}$ which diagonalizes the BdG Hamiltonian,
\begin{equation}
    H = \frac{1}{2} \begin{pmatrix} c^\dagger & c \end{pmatrix} U U^\dagger \mathcal{H}_{\textrm{BdG}} U U^\dagger \begin{pmatrix} c \\ c^\dagger \end{pmatrix} = \frac{1}{2} \begin{pmatrix} \gamma^\dagger & \gamma \end{pmatrix} \mathcal{H}_{\textrm{diag}} \begin{pmatrix} \gamma \\ \gamma^\dagger \end{pmatrix},
\end{equation}
with $c = \begin{pmatrix}c_1 & c_2 & ... \end{pmatrix}^T$. For the n-th eigenmode, the Bogoliubov operator $\gamma(E_n)$ is written as
\begin{equation} \label{bogoliubons}
\begin{split}
    \gamma (E_n) &= \phi_n^\dagger (E_n) \begin{pmatrix} c \\ c^\dagger \end{pmatrix} = \begin{pmatrix}u_n & v_n \end{pmatrix} \begin{pmatrix} c \\ c^\dagger \end{pmatrix} \\
    &= \sum_j \left(u_{j, n} c_j + v_{j,n} c^\dagger_j \right).
\end{split}
\end{equation}

Due to particle-hole symmetry, each state $\phi_n(E_n)$ has a particle-hole partner $\mathcal{P} \phi_n(E_n) = \phi_{n}(-E_n)$, with energy $-E_n$, where $\mathcal{P} = \tau_x \mathcal{K} \otimes \mathds{1}_{N}$ is the particle-hole operator. Here $\tau_x$ is a Pauli matrix that flips the electron and hole components, and $\mathcal{K}$ is the complex conjugation operator. For each state $\mathcal{P}\phi_n(E_n) = \phi_n (-E_n)$, we have the following operator

\begin{equation} \label{particle hole partner operator}
    \gamma (-E_n) = \phi_n^\dagger (-E_n) \begin{pmatrix} c \\ c^\dagger \end{pmatrix}  = \sum_j \left(v_{j, n}^* c_j + u_{j,n}^* c^\dagger_j \right) = \gamma^\dagger (E_n).
\end{equation}
For a Majorana operator, $\gamma(E_n)  = \gamma^\dagger (E_n)$, meaning that $\mathcal{P} \phi_n (E_n) = \phi_n (-E_n) = \phi_n (E_n)$, a condition that can hold only for $E_n = 0$.

When the energy of the lowest mode is zero, the nature of the operator shifts from a fermionic to a Majorana operator. This distinction becomes clear by calculating $\gamma^2(E_n)$: for fermionic operators, which obey Pauli's principle, $\gamma^2(E_n \neq 0) = 0$. In contrast, Majorana operators are characterized by $\gamma^2 (E_n = 0) = 1/2$. In terms of Bogoliubov coefficients ($u_j$, $v_j$), we have

\begin{equation} \label{gamma2 expression}
    \gamma^2(E_n) = \sum_j u_{j, n} v_{j,n}.
\end{equation}
Numerically, since the energy is never zero, we identify a true zero-energy mode by the self-adjointness of the $\gamma(E_M)$ operator, indicated by $\gamma^2 = 1/2$. In contrast, finite-energy fermionic excitations are characterized by $\gamma^2 = 0$. Additionally, we find that $\bra{\phi_n} \mathcal{P}\ket{\phi_n} = 2\sum_j u_{j, n} v_{j, n} = 2 \gamma^2 (E_n)$, confirming that PMMs/MBSs represent eigenstates of the particle-hole operator. This arises because the associated wave function exhibits equal contributions from the electron and hole parts.

\subsection{Majorana Polarization}

In long Kitaev chains, the overlap between the end MBSs is negligible when the localization length is much smaller than the chain length. This is the desirable configuration for operations that exploit the non-Abelian nature of these MBSs/PMMs, for instance, braiding operations \cite{aliceaDephasing}. In short chains, however, even small deviations from the sweet spot may lead to significant overlap between the PMMs. Here we quantify the spatial overlap of the PMMs/MBSs using the concept of MP introduced in Ref. \cite{sedlmayr2015visualizing}, which we present below.

We start by decomposing the fermionic operators of the Kitaev chain as
\begin{equation}
	\label{c in majoranas}
    c_j= \frac{\gamma_{B, j} + i\gamma_{A, j}}{\sqrt{2}},
\end{equation}
where,
\begin{equation}
    \gamma_{A, j} = \frac{i (c_j^\dagger - c_j)}{\sqrt{2}}, \,\,\, \gamma_{B, j} = \frac{c_j^\dagger + c_j}{\sqrt{2}}.
\end{equation}
We now substitute \eqref{c in majoranas} into the solutions given by  Eq.~\eqref{bogoliubons}. The Bogoliubons in terms of Majorana operators are given by
\begin{equation} \label{gamma in majoranas}
\begin{split}
\gamma (E_n) =& \psi_n^\dagger (E_n) \begin{pmatrix}
    \gamma_A \\ \gamma_B
\end{pmatrix} = \begin{pmatrix}
    i \psi_A & \psi_B
\end{pmatrix} \begin{pmatrix}
    \gamma_A \\ \gamma_B
\end{pmatrix} \\
&= \sum_j i a_{j, n} \gamma_{A, j} + b_{j, n} \gamma_{B, j},
\end{split}
\end{equation}
with $\psi_A = \begin{pmatrix}
    a_{1, n} & a_{2, n} & ... & a_{N, n}
\end{pmatrix}$, $\psi_B = \begin{pmatrix}
    b_{1, n} & b_{2, n} & ... & b_{N, n}
\end{pmatrix}$, $a_{j, n} = \frac{u_{j, n} - v_{j, n}}{\sqrt{2}}$ and $b_j = \frac{u_{j, n} + v_{j, n}}{\sqrt{2}}$. Henceforward, we focus on the lowest-energy mode and omit the $n$ index. 

The distinction between MBSs/PMMs and trivial Andreev bound states lies in the spatial localization of single MBSs at the opposite ends of the chain. This separation implies no overlap between the Majorana wave functions $\psi_A$ and $\psi_B$. To quantify the isolation of MBSs at the ends of the chain we use the MP, defined as

\begin{equation} \label{MP definition}
    M_j = \frac{b_j^2 - a_j^2}{b_j^2 + a_j^2} = \frac{2 u_j v_j}{u_j^2 + v_j^2}.
\end{equation}

Note that $M_j$ also gauges the robustness of the zero-energy modes against local perturbations. This is expressed through the product of $u_j$ and $v_j$ in Eq. (\ref{MP definition}), which in turn reflects the particle-hole operator's projection at a given site,
$\bra{\phi_{j,n}}\mathcal{P} \ket{\phi_{j,n}} = 2 u_j v_j$. This expectation value is maximized when $|u_j| = |v_j|$. In this case, either $a_j$ or $b_j$ vanishes and there is no overlap between $\psi_A$ and $\psi_B$. Hence, $|M_j| = 1$, signaling that a 
local perturbation is unable to cause an energy-splitting because the local charge, $q_j = -e \left(u_j^2 - v_j^2 \right)$ ($e = |e|$) \cite{flensbergBCSCharge}, is zero.  Conversely, for an electron or a hole state, either $u_j$ or $v_j$ vanishes ($|a_j| = |b_j|$), yielding $|M_j| = 0$, which indicates full overlap of the Majorana wave functions on this site. This configuration is vulnerable to electrical impurities, as the local BCS charge $q_j$ is maximized. In summary, while states can always be represented in a Majorana basis, they correspond to PMMs or MBSs only when their wave functions are spatially separated, as indicated by high values of the MP.

We observe a connection between the expressions for $\gamma^2$ and $M_j$, Eqs. (\ref{gamma2 expression}) and (\ref{MP definition}, respectively, in terms of the product of the Bogoliubov coefficients. The expression for $\gamma^2$ probes the non-Abelian nature of the whole wave function, as $\gamma^2 = 1/2$ signals an exact zero-energy eigenstate of the particle-hole operator. On the other hand, the MP provides information about the non-Abelian nature of the excitation in a given region of space by quantifying how well-separated the Majorana wave functions are. In combination, these two quantities gauge how much the solutions satisfy the particle-hole operator on a local and global scales, which directly translates into an assessment of the robustness against perturbations. For this reason, we characterize the sweet spots based on the degeneracy of the ground state, here given by $E_M = 0$, and $|M| = 1$ at the edges of the Kitaev chain, where the Majorana wave functions localize.

\subsection{2-site Kitaev chain}

We close this section by applying the above concepts to the 2-site Kitaev chain. By diagonalizing the BdG Hamiltonian, we find the eigenfunction for the lowest-energy mode ($\Delta = 1$, $\mu_1 = \mu_2$),

\begin{equation}
    \phi = \frac{1}{\mathcal{C}} \begin{pmatrix}
    -\mu - \sqrt{1 + \mu^2} & \mu + \sqrt{1 + \mu^2} & 1 & 1
\end{pmatrix}^T,
\end{equation}
where $\mathcal{C}^2 = 4 \sqrt{1 + \mu^2}\left( \mu + \sqrt{1 + \mu^2} \right)$. Using Eq. (\ref{MP definition}), we obtain the MP

\begin{equation}
    |M| = \frac{1}{\sqrt{1 + \mu^2}},
\end{equation}
where $|M_1| = |M_2| = |M|$. From the above equation, we observe that the MP is maximized $|M| = 1$ for $\mu = 0$. As discussed above, this is associated with the local BCS charges $q_1$ and $q_2$, given by

\begin{equation}
    q_1 = q_2 =  \frac{-e \mu}{2 \sqrt{1 + \mu^2}},
\end{equation}
i.e., for $\mu = 0$ the local charges vanish, which is associated with a peak in the MP.

Taking $\mu = 0$, the eigenenergies of the BdG Hamiltonian are

\begin{equation}
    E_M = |1 - t| , \,\,\,\, E_1 = 1 + t.
\end{equation}
Therefore, we note that for $\mu = 0$ and $t = \Delta = 1$, $E_M = 0$ and $|M| = 1$, which fully characterizes the 2-site sweet spot \cite{leijnsepoorman}. 

Finally, the operator associated with the lowest-energy mode naturally becomes a Majorana operator when $E_M = 0$. To make a clear distinction, we first consider $t \neq \Delta$. The eigenvectors associated with $\pm E_M$ read

\begin{equation}
    \phi = \frac{1}{2} \begin{pmatrix}
    1 & 1 & -1 & 1
    \end{pmatrix}^T, \,\,\,\, \mathcal{P} \phi = \frac{1}{2} \begin{pmatrix}
    -1 & 1 & 1 & 1
    \end{pmatrix}^T.
\end{equation}
We note that $\gamma^2(E_M) = \sum_i u_i v_i = [1 (-1) + 1 (1)]/2 = 0$, as expected for a non-zero energy level. Also, it naturally follows that $\bra{\phi}\mathcal{P}\ket{\phi} = 0$, as $E_M \neq 0$ implies that $\phi$ and $\mathcal{P}\phi$ are orthogonal. Now we consider $t = \Delta$, such that $E_M = 0$. In this case, $\mathcal{H}_{BdG} \psi = 0$ and the eigenvectors are

\begin{equation}
    \phi_A = \frac{i}{\sqrt{2}} \begin{pmatrix}
    1 & 0 & -1 & 0
    \end{pmatrix}^T, \,\,\,\, \phi_B = \frac{1}{\sqrt{2}}\begin{pmatrix}
    0 & 1 & 0 & 1
    \end{pmatrix}^T.
\end{equation}
Note that the two eigenvectors are now eigenstates of the particle-hole operator with, $\mathcal{P} \phi_{A, B} = \phi_{A, B}$. According to our previous discussion, the operators associated with these eigenstates are Majorana operators. From the above eigenstates, $\gamma_A^2 = \gamma_B^2 = 1/2$.

\section{Scaling short Kitaev chains} \label{scaling short chains}

Now that we have characterized the sweet spots based on the ground state degeneracy and MP requirements, we investigate the evolution of the 2-site sweet spot as we increase the length of the Kitaev chain. To this end, we calculate $E_M$ and $|M|$ for an increasing number of sites. 

In the first column of Fig.~\ref{Fig1}, we observe that as $N$ increases, the white region ($E_M \to 0$) around the sweet spot ($\mu = 0$, $t = \Delta$) grows because more zero-energy lines are crossing that point. As mentioned previously, there is a new line for each added site in the chain, see Eq. (\ref{majoranaLines}). The lines are symmetric for positive and negative values of $\mu$, such that for odd-$N$, $\mu = 0$ will always provide an exact zero-energy solution. Since $E_M$ decreases closer to the solutions given by Eq. (\ref{majoranaLines}), the convergence of these lines at $\mu = 0$ and $t = \Delta$ significantly increases the white regions in the first column of Fig. \ref{Fig1} as we go from $N = 2$ to $N = 5$ and $9$.

Next, we focus on the MP, whose results we show in the second column of Fig. \ref{Fig1}. Interestingly, there are $N-1$ lines in parameter space that characterize $|M_1| = |M_N| = |M| = 1$. Similar to the $E_M = 0$ lines in the first column, the $|M| = 1$ solutions also converge to the 2-site sweet spot. In addition, the value of the MP decays monotonically as one moves away from the white lines, which means that the increasing number of (sites) lines crossing the 2-site sweet spot gives rise to a stable high MP region. Therefore, while there was initially ($N = 2$) a single point where we could find an intersection between $E_M = 0$ and $|M| = 1$, for $N \geq 5$, there is now an entire region. Essentially, the fine-tuning conditions required to demonstrate PMMs in 2-site Kitaev chains are relaxed once the number of sites increases, as the sweet spot grows into a ``sweet region", whose area is defined by the number of sites. 

Hence as $N$ increases, the PMMs become more robust against parameter variation, as depicted by the white regions shown in Fig. \ref{Fig1}. In general, for a Kitaev sweet spot with $N$ sites, up to $N-1$ chemical potentials can be varied while keeping $E_M = 0$ \cite{bordin2024signatures}. Here, we demonstrate that for $N \geq 5$, the energy remains near zero, and the MP stays close to 1, even when all $N$ chemical potentials and hoppings are varied within the white regions. Our findings align with recent experimental observations. For instance, Ref.~\cite{bordin2024signatures} reports that increasing the number of sites in artificial Kitaev chains from $N = 2$ to $N = 3$ significantly enhances the robustness of the zero modes. Similarly, Ref.~\cite{Haaf_arXiv2024} demonstrates the transition from $N = 2$ to $N = 3$ and the corresponding emergence of an additional zero-energy line, as illustrated in Figs.~\ref{Fig1}(a, b).

We note that although small in comparison to $\Delta$, $E_M$ remains finite for most of the white regions in Fig. \ref{Fig1}, which corresponds to fermionic excitations with $\gamma^2 = 0$. This implies that if we select the parameters given by Eq. (\ref{majoranaLines}) where $E_M = 0$, even small fluctuations might cause an energy splitting in the spectrum. Thus, we consider the Majorana modes in this limit ($N < 10$) as PMMs. For longer chains $N > 20$, we identify regions exhibiting topological protection, where the ground state remains degenerate even in the presence of disorder. This indicates that the self-adjointness of the excitations is preserved despite parameter fluctuations, see Sec. (\ref{long chains}). Therefore, the transition between short ($N<10$) and long ($N>20$) Kitaev chains represents a crossover between unprotected PMMs and MBSs.

\section{Spinful model at finite Zeeman field} \label{spinfulmodel Section}

\begin{figure}
	\centering	\includegraphics[width=0.475\textwidth]{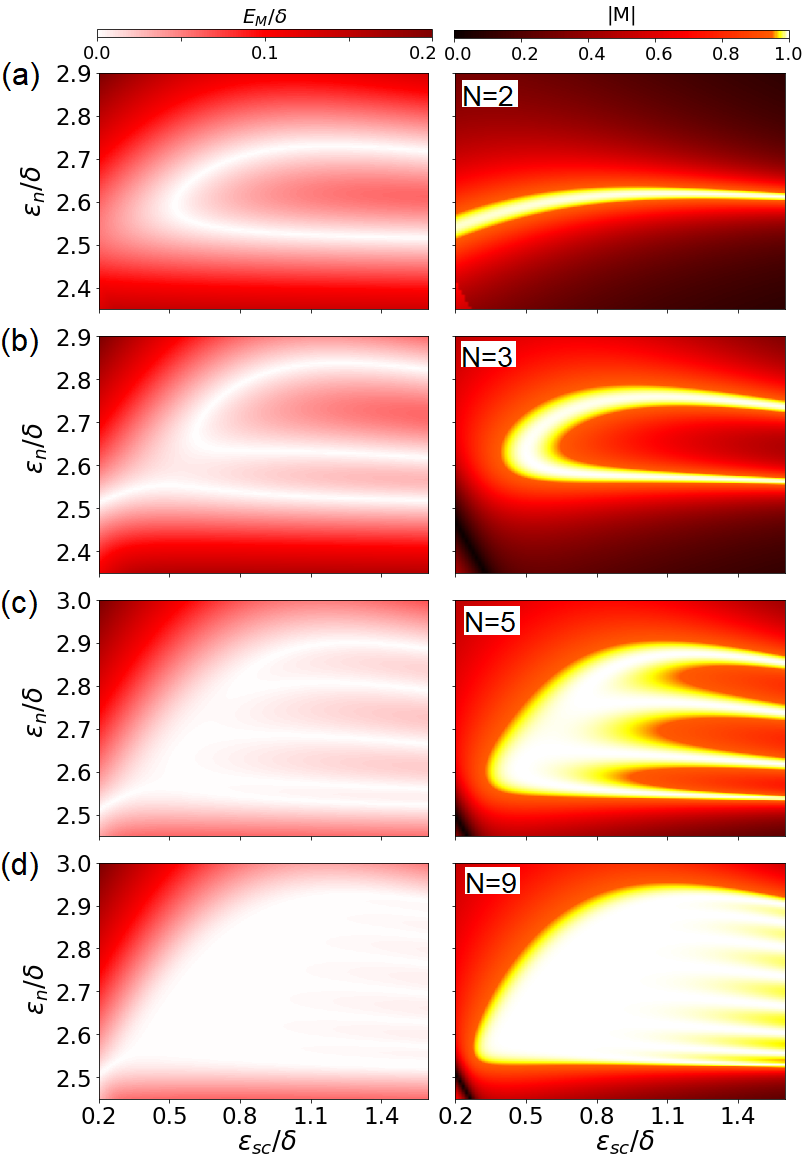}
	\caption{Energy of the first excited state $E_M$ (left panel) and MP (right panel) as functions of the level of the normal $\varepsilon_n$ and superconducting $\varepsilon_{sc}$ QDs. (a)-(d) $N=2, 3, 5, 9$, respectively, where $N$ is the number of normal dots. Similar to the Kitaev chain, Fig. \ref{Fig1}, the convergence of zero-energy and $|M| = 1$ lines expands the sweet spot into a region, whose area increases with $N$.}
\label{Fig2}
\end{figure}

We now investigate the evolution of the sweet spot for the more realistic model, which considers both spin channels and finite magnetic fields, as described by Eq. (\ref{Hspinful}). To cast the Hamiltonian into the BdG form, we have neglected Coulomb interaction terms, which become more relevant in the regime of weak magnetic fields ($V_z < \delta$) \cite{tsintzis2022creating}. Here, we set $V_z = 2.5\delta$, $t = 0.5\delta$, and $t^{so} = 0.2t$ \cite{tsintzis2022creating}. Due to screening effects, we consider $V_z = 0$ in the superconductors, but finite values do not change our conclusions \footnote{We note that the magnetic field can not be arbitrarily increased in the superconductors as it might suppress the s-wave pairing amplitude $\Delta$}. We show the results for $E_M$ and $|M|$ ($|M_j|=|M|$ for $j=1$ and $j=N$) as functions of the gates applied to the normal ($\varepsilon_{n}$) and superconducting ($\varepsilon_{sc}$) QDs in Fig. (\ref{Fig2}).

We first focus on the 2-site Kitaev chain, Fig.~\ref{Fig2}(a). Here, the two normal QDs are coupled via discrete ABS levels, tuned via $\varepsilon_{sc}$. Due to the spin-orbit coupling and the applied magnetic field, the normal QDs can be connected via same-spin CAR. In addition, electrons can tunnel between the normal QDs via ECT. If we then project the superconducting dot out, we obtain an effective 2-site Kitaev chain, Eq.~(\ref{KitaevChain}), where the p-wave pairing $\Delta$ is proportional to the CAR amplitude and the hopping $t$ to the ECT. As discussed previously, to achieve a sweet spot we need to tune the parameters such that $t = \Delta$, which is done by varying $\varepsilon_{sc}$. The second requirement is to set the chemical potential of the QDs such that $\mu_i = 0$ in the effective Kitaev chain. For this reason, we also vary $\varepsilon_n$. The sweet spot is characterized by the intersection between the white lines on the left ($E_M = 0$) and right ($|M| = 1$) panels in Fig.~\ref{Fig2}(a). Analogously to the Kitaev model, Fig.~\ref{Fig1}(a), we have 2 lines that describe a family of zero-energy solutions. Additionally, we observe a white line on the right panel that describes $|M| \to 1$. Differently from the Kitaev model, though, for finite magnetic fields, it is seldom possible to observe fully separated PMMs \cite{tsintzis2022creating, PhysRevB.110.245412} in 2-site sweet spots, here we observe a maximum of $|M| \approx 0.99$. 

We now increase the number $N$ of normal QDs, which translates into adding more sites to the effective Kitaev chain. In Fig. \ref{Fig2}(b), we observe that the energy and MP plots present an additional line, emulating a 3-site Kitaev chain in this case. Here, we note an avoided crossing in the zero-energy lines due to the emergence of next-to-nearest-neighbor couplings in the effective Kitaev chain \cite{svensson2024quantum, miles2024kitaev}. This level repulsion asymptotically goes away as we increase the magnetic field, fully polarizing the system. By further increasing $N$, we note that the number of zero-energy lines perfectly matches the ones described by Eq. (\ref{majoranaLines}). Furthermore, the intersection of lines at the initially 2-site sweet spot creates a robust white region where $E_M \to 0$. 

The number of $|M| \to 1$ lines also increases for each additional normal QD, similarly to the results in Sec. (\ref{scaling short chains}). This is shown in the right column of Fig. \ref{Fig2}, where the number of lines is always $N - 1$. For $N = 2$, the maximum value of $|M|$ depends on $V_z$, as stronger (weaker) magnetic fields provide a better (worse) localization of the PMMs in the normal dots. For the same value of $V_z$, we notice that by adding QDs to the chain, the value of MP increases, see the emergence of white lines for $N \geq 3$. More importantly, the convergence of the $|M| \to 1$ lines near the original 2-site sweet spot creates a robust white region that overlaps with the $E_M \to 0$ region, see Fig. \ref{Fig2}(d) left panel. 

Here, by comparing Figs. \ref{Fig2}(c) and \ref{Fig2}(d) we note the emergence of a well-delimited white region that will eventually become a topological island. This occurs because additional lines will be enclosed by the borders shown for $N = 9$. This observation is also aligned with the results for the Kitaev chain, see Fig. \ref{Fig1}(d), where we see the borders of the white region at $|\mu| = 2t$, which signals the topological phase transition for $N \to \infty$. 

In the realistic model, unlike in the Kitaev model, the convergence point for the zero-energy solutions is not exactly the 2-site sweet spot. This occurs because in the transition from $N = 2$ to $N > 2$, normal QDs in the middle of the chain couple to two superconducting dots, whereas the outermost QDs (left and right) couple to only one. These different couplings lead to distinct renormalizations of the parameters that affect the ECT and CAR amplitutes. It is then necessary to adjust the QD levels, which in turn shift the sweet spots. For certain parameters, the original 2-site sweet spot might lie outside the topological regions identified for large $N$. This fact has no impact on the essential properties of the 2-site sweet spots, such as non-Abelian exchange statistics, zero-energy, and strong localization. However, it is important to note that when gradually increasing the number of QDs in the chain, one must broaden the parameter space window to locate the convergence of the zero-energy lines.

\section{Long Kitaev chains - Zero-energy plateaus} \label{long chains}

We now present the results for longer Kitaev chains $N \geq 20$. The goal here is to sweep the parameter space near the 2-site sweet spot and verify whether there are regions within which the operators follow non-Abelian statistics, i.e., there is no hybridization energy $E_M = 0$ associated with the MBSs, see Eq. (\ref{gamma2 expression}). We also include fluctuations in the parameters to test the robustness of the results.

\begin{figure}
	\centering
	\includegraphics[width=0.475\textwidth]{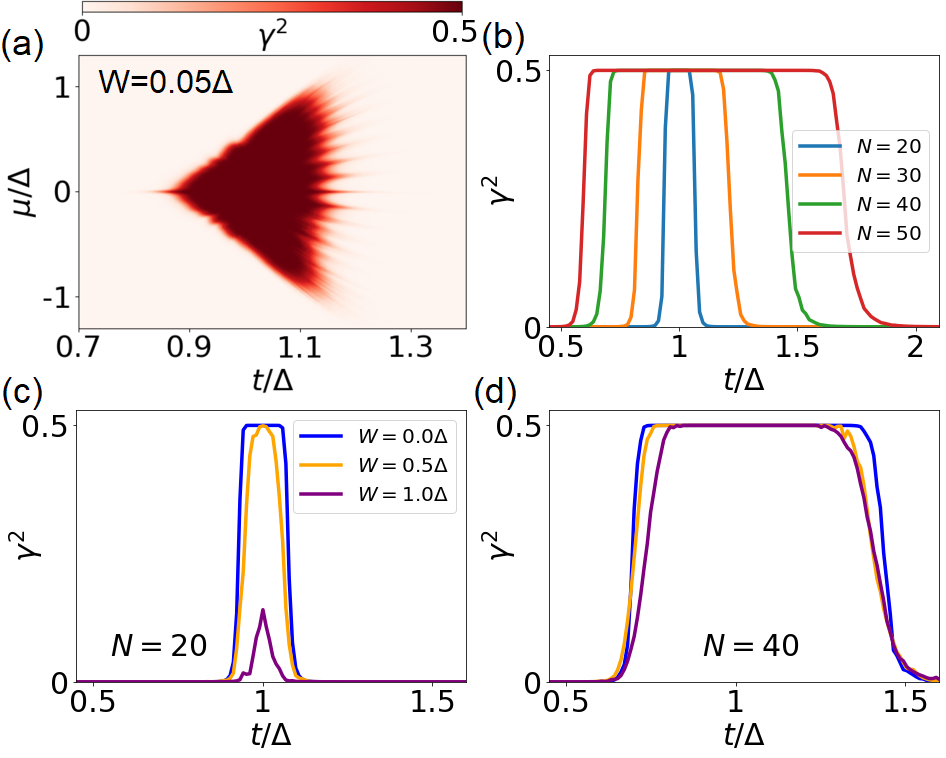}
	\caption{(a) $\gamma^2$ as a function of $\mu/\Delta$ and $t/\Delta$ for $N = 25$ with random fluctuations in the hopping and chemical potential of $5\%$ of $\Delta$. We average the results over $50$ disorder realizations. We observe the emergence of a triangular region where $\gamma^2 = 1/2$, which can be reproduced (with limited precision due to finite temperature) by conductance measurements, as shown in Figs.~\ref{Fig6}(a-b). (b) $\gamma^2$ as a function of $t/\Delta$ for different values of $N$, showing that the zero-energy plateau gets significantly wider when N increases. (c)-(d) Robustness of the zero-energy plateau against disorder for (c) $N = 20$ and (d) $N = 40$, averaged over $100$ disorder realizations for $W = 0, 0.5\Delta$, and $\Delta$, represented in the plots by the blue, orange, and purple lines, respectively. (c) The shorter chain presents smaller plateaus that vanish as the disorder strength increases. (d) The longer chain shows larger plateaus which are stable in the presence of disorder. For (b)-(d), $\mu = 0$.}
	\label{Fig4}
\end{figure}

The high density of $E_M = 0$ lines near the sweet spot gives rise to strictly zero-energy plateaus in parameter space, which scale with the number of sites. To show this, we first calculate $\gamma^2$ for the lowest-energy mode, according to Eq. (\ref{gamma2 expression}), for $N = 25$ and vary $\mu/\Delta$ and $t/\Delta$. We also include random fluctuations of $5 \%$ of the gap in all parameters and average the results over $50$ realizations. The results are shown in Fig. \ref{Fig4}(a), where we observe a region with $\gamma^2 = 1/2$ near $\mu = 0$ and $t = \Delta$. As discussed in Sec. (\ref{degeneracy and MP}), a self-adjoint operator $\gamma^2 = 1/2$ implies an exact zero-energy mode due to particle-hole symmetry. Away from the 2-site sweet spot, we note that fluctuations lift the ground state degeneracy along the lines described by Eq. (\ref{majoranaLines}). However, the energy in the region surrounding this sweet spot, where the lines converge, remains stable at zero. Due to the protection of the MBSs, we identify the triangular area in Fig. \ref{Fig4}(a) as a topological island. Thus we observe the evolution of the 2-site Kitaev chain sweet spot into a low-energy, high-MP region for short chains, and ultimately into a topological island as the chain length increases, as shown in Figs. \ref{Fig1} and \ref{Fig4}(a), respectively. We verify that the MP is also maximized within the topological island ($|M| > 0.9999$). 

To show the increase in the topological region depicted in Fig. \ref{Fig4}(a) with the number of sites, we plot $\gamma^2$ as a function of $t$ for $\mu = 0$ and $W = 0$ for several values of $N$ in Fig.~\ref{Fig4}(b). As anticipated, the width of the plateau significantly increases with $N$. For larger values of
$|\mu|$, its extent diminishes, as indicated in Fig.~\ref{Fig4}(a). In the limit $N \to \infty$, the entire parameter space for $t > |\mu|/2$ is filled with the solutions of Eq.~(\ref{majoranaLines}), such that $E_M = 0$ and $\gamma^2 = 1/2$. At this point, the 2-site sweet spot, which has become the triangular region in Fig.~\ref{Fig4}(a), has evolved into the entire topological phase. 

Long chains present remarkable robustness against disorder in the form of stable $\gamma^2 = 1/2$ plateaus. As the chain length increases, the hybridization decreases, resulting in wider plateaus. Additionally, the width of these plateaus reflects the system's stability against parameter fluctuations. To understand this result, we consider a Jackiw-Rebbi-type problem \cite{jackiw1976solitons}. For a semi-infinite Kitaev chain in the continuous limit ($k \to 0$) and the chemical potential described by $\mu(x) = \mu \,  \textrm{sgn}(x)$, the domain wall at $x = 0$ causes the emergence of a single MBS. Notably, the energy of this MBS is pinned to zero as long as particle-hole symmetry is not broken. Thus, for finite chains, as we approach the infinite chain limit by gradually increasing $N$, we expect the zero-modes to become more resilient against local (Anderson) disorder. We confirm this assumption with numerical simulations. In Figs.~\ref{Fig4}(c-d), we calculate $\gamma^2$ as a function of $t/\Delta$ for increasing disorder strengths, $W = 0$, $0.5\Delta$, and $ \Delta$ (blue, orange, and purple lines, respectively) and different chain sizes, $N = 20$ (c) and $40$ (d). We average the results over $100$ disorder realizations. For shorter chains, we observe that for $W = 0.5\Delta$ the plateau is already significantly depleted, and no plateau exists for $W = \Delta$. In contrast, the longer chain shows remarkable robustness against disorder, as $W = 0.5\Delta$ leaves the plateau almost unchanged, and even for $W = \Delta$ there is a reliable zero-energy region around $t = \Delta $.

The zero-energy plateaus might be helpful in terms of building Majorana-based qubits in longer chains of QDs. For instance, it was verified in Ref. \cite{aliceaDephasing} that the dephasing times of these qubits rely on the derivative of the energy splitting with respect to fluctuations in the parameters, which vanishes within the topological islands. In this sense, not only do we avoid dynamical phases by setting $E_M = 0$, but we also show a way of increasing the coherence times. 

Finally, another important aspect of realistic setups is the possibility of coexistence between MBSs with low-energy fermionic excitations. It was shown in Ref. \cite{akhmerov2010topological} that this type of interaction does not affect the topological protection nor the non-Abelian braiding rules of the Majorana operators as long as $E_M = 0$, again emphasizing the importance of the topological islands described in this section.

\section{Conductance in the QD-Kitaev setup} \label{conductance section}

\begin{figure}
	\centering
\includegraphics[width=0.475\textwidth]{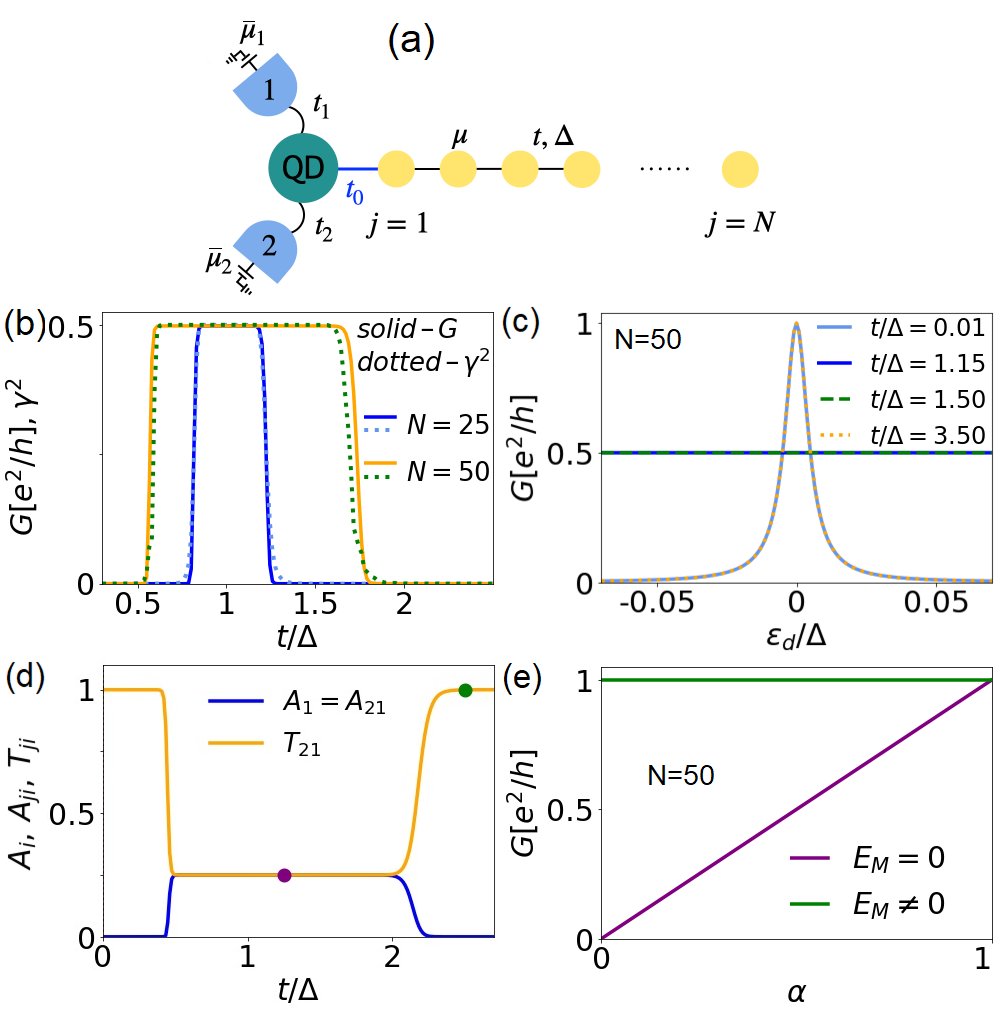}
	\caption{(a) Schematic of a Kitaev chain with an arbitrary number of sites side-coupled to a probe QD. (b) Conductance and $\gamma^2$ as a function of $t/\Delta$. The solid and dotted lines represent the conductance, in units of $e^2/h$, and $\gamma^2$, respectively. The matching between solid and dotted curves, obtained by setting $\varepsilon_d/\Delta = -10$, enables the conductance to probe the topological islands. (c) Conductance as a function of $\varepsilon_d$. By setting the parameters inside the zero-energy plateau (blue line), the conductance is mediated by the Majorana mode that leaks into the QD and is independent of the dot level. Outside the plateau (light-blue and orange lines), the Majorana mode does not contribute to the conductance, which depends solely on the dot level. In this case, the conductance shows a Lorentzian-shaped dependence with $\varepsilon_d$. (d) Probability coefficients for direct transport, $T_{21}$, and both local and crossed AR, $A_1 = A_{21}$, as a function of $t/\Delta$. Inside the Majorana plateau, $T_{21} = A_{21} = A_1 = 1/4$. (e) G as a function of $\alpha$. Since Andreev processes depend on the voltage drop configuration, the conductance depends linearly on $\alpha$ (purple curve). Outside the plateau, the resonant dot level ensures $T_{21} = 1$, making $G$ independent of $\alpha$ (green curve). The parameters used are $\mu/\Delta = 0.001$, $N = 25$ [(b) only] and $50$, $\Gamma_1 = \Gamma_2 =  0.005 \Delta$, and $t_0 = 0.02 \Gamma_1$.}
	\label{Fig5}
\end{figure}

After investigating the different regimes for the Kitaev model, we study the conductance associated with the transversal current flowing through the probe-QD in the system described by Eqs. \eqref{FullHdotKitaev}-\eqref{leadsDot}, and sketched in Fig. \ref{Fig5}(a). We recover the findings of \cite{barangerDot, vernekLeakage} and generalize the results to finite chains. Furthermore, we reinterpret the conductance by detailing the different processes involved. Here, the main goal is to exploit the tunable QD level to improve the resolution in spectroscopic measurements and associate peaks in the conductance with the algebra of the Bogoliubov operators.

The conductance is calculated by injecting electrons in one of the leads and calculating the reflection and transmission probability amplitudes. As we shall see, the incoming electron can tunnel to the other lead as an electron or as a hole, characterizing the ECT and CAR processes. In addition, the electron can be reflected as a hole in the same lead, a process known as local Andreev reflection. To obtain the reflection and transmission coefficients, we calculate the scattering matrix using the general expression \cite{splittingBeenakker, reproducingAkhmerov, dresen2014quantum}
\begin{equation} \label{smatrixExpression}
	S(E) = \mathds{1} - 2 \pi i \rho  W^\dagger(E + i \pi \rho W W^\dagger - \mathcal{H}_0)^{-1}W,
\end{equation}
where $\rho$ is the density of states (DOS) of the leads, $\mathcal{H}_0$ corresponds to the BdG Hamiltonian of Eq. (\ref{dotKitaevH}), and $W$ couples the system with the leads, see App. (\ref{App A}). The s-matrix provides the coefficients that allow us to calculate the current through the probe-QD \cite{btk, flensbergBCSCharge},

\begin{equation} \label{I1}
	\begin{split}
		I_1 &= \frac{e}{h} \int dE \left(2 A_1 + T_{21} + A_{21} \right) \Tilde{f}(\Bar{\mu}_1) \\
		&- \frac{e}{h} \int dE \left(T_{12} - A_{12} \right) \Tilde{f}(\Bar{\mu}_2),
	\end{split}
\end{equation}
where $A_1$ and $A_{21}$ represent the local and crossed Andreev reflection processes, respectively, and $T_{21}$ denotes the electron tunneling. Also, $\Tilde{f}(\Bar{\mu}_i) = f(E - \Bar{\mu}_i) - f(E)$, and $f(E) = \left[1 + e^{(E/k_B T)} \right]^{-1}$ is the Fermi function. We consider a generic voltage drop between leads $1$ and $2$, see Fig. \ref{Fig5}(a), $V_1 = \alpha V$, $V_2 = ( \alpha - 1) V$, such that $V_1 - V_2 = V$, with $\Bar{\mu}_{1,2} = e V_{1,2}$ measured with respect to the grounded superconductor \cite{flensberg2022conductance}. Finally, we define the conductance as $G = dI_1/dV$.

To obtain analytical expressions, we first focus on the regime of low energies and linear response at $T=0$. The detailed derivation is carried out in App. \ref{App A}. To simplify the equations, we substitute $T_{12}(E) = T_{21} (E)$, which is ensured by the unitarity of the s-matrix in one-dimensional systems \cite{lesovik2011scattering}. In addition, we find that for $\Gamma_1 = 2 \pi \rho t_1^2  = \Gamma_2 = \Gamma$, $A_1 (E) = A_{21} (E)$ which means that the probability of the electron being transferred or reflected as a hole is the same for this setup. This relation does not come from a symmetry of the Hamiltonian, but from the matrix $W$ because the leads are connected to the same site. Finally, we assume that the tunneling rates $\Gamma_i$ are independent of the energy (wide-band limit). After these considerations, the conductance assumes the compact form

\begin{equation} \label{GSimpleForm}
	G = \frac{e^2}{h}\left[ A_1 (4\alpha - 1) + T_{21}\right]_{E \to 0}.
\end{equation}

When there is a hybridization between the MBSs ($E_M \neq 0$), the conductance is mediated solely by the level of the probe-QD, with no contribution from the Kitaev chain. To show this, let us consider $E_M \neq 0$. For $E = 0$, we obtain $A_1 = 0$ and $T_{21} = \Gamma^2 |A_{11}|^2$, see App. \ref{App A}. In this case, the conductance is given by
\begin{equation} \label{dotConductance}
    G = \frac{e^2}{h}\frac{\Gamma^2}{\Gamma^2 + \varepsilon_d^2}.
\end{equation}
For $\varepsilon_d = 0$, $T_{21} = 1$ and $G = e^2/h$. Interestingly, this is the usual conductance of a QD ($t_0 = 0$) coupled to metallic leads. This result shows that for a finite Kitaev chain, the conductance is unaffected by the MBS when $E_M \neq 0$. On the other hand, if $E_M = 0$, we find that $A_{1} = T_{21} = 1/4$. The conductance reads

\begin{equation} \label{majoranaConductance}
    G = \frac{\alpha e^2}{h}.
\end{equation}
The emergence of Andreev processes at $E = 0$ makes the conductance dependent on the voltage drop configuration. In particular, a common choice is to use a symmetric voltage drop, $\alpha = 1/2$, as it ensures current conservation. In this case, we recover the results of \cite{vernekLeakage} with $G = e^2/2h$ for infinitely long chains.

Now we provide a clear relation between the algebra of the Bogoliubov operator associated with the lowest-energy mode, Eq. (\ref{bogoliubons}), and the zero-bias conductance,

\begin{equation}
    G = \begin{cases}
			0, & \text{for } \gamma^2 = 0, \\
            e^2/2h, & \text{for } \gamma^2 = 1/2.
		 \end{cases}
\end{equation}

We compare these analytical results with numerical simulations using the full model, see Eq. (\ref{FullHdotKitaev}). In Fig. \ref{Fig5}(b), we set $\varepsilon_d = -10\Delta$ and calculate the conductance (solid lines) and $\gamma^2$ (dotted lines) as functions of $t$. We observe that for both $N = 25$ and $50$, the conductance vanishes when $\gamma^2 = 0$, and peaks at $e^2/2h$ when $\gamma^2 = 1/2$. Therefore, the conductance in the probe-QD can directly track the zero-energy plateaus in this limit.

Interestingly, when $E_M = 0$, $G$ does not depend on the energy of the probe-QD, which gives rise to a conductance plateau with respect to $\varepsilon_d$, cf. Ref. \cite{vernekLeakage}. This result is corroborated by a numerical simulation, shown in~Fig. \ref{Fig5}(c), where we plot $G$ as a function of $\varepsilon_d$ for several values of $t$. For values of $t$ outside the plateau in Fig.~\ref{Fig5}(b), $E_M \neq 0$ and the conductance reproduces the Lorentzian shape of Eq. (\ref{dotConductance}), see the solid light blue and dotted orange lines. Conversely, if we choose $t$ to be inside the zero-energy plateau, the conductance is unaffected by $\varepsilon_d$, as shown by the solid blue and dashed green curves.

An alternative way of probing zero-energy solutions is by exploring the dependence of the conductance on the voltage drop configuration, $\alpha$. In Fig. \ref{Fig5}(d), we set $\varepsilon_d = 0$ and plot the electron tunneling $T_{21}$ and the local Andreev Reflection $A_1$ coefficients as functions of t. As discussed above, when $E_M = 0$, $A_1 = T_{21}$, which characterizes the topological island. Outside of the limits of the zero-energy plateau, the Andreev reflection vanishes and the electron tunneling goes from $1/4$ to $1$. In comparison to Fig. \ref{Fig5}(b), the plateau has slightly increased due to the coupling of hybridized MBSs with the probe-QD level. As anticipated by Eq. (\ref{GSimpleForm}), when $A_1 \neq 0$, the conductance depends on the voltage bias configuration. We select a point (purple dot) within the topological island to show this, see the purple curve in Fig. \ref{Fig5}(e). For $E_M \neq 0$, $A_1 = 0$ and $G$ is independent of $\alpha$, as shown by the green line in Fig. \ref{Fig5}(e). As we increase $\varepsilon_d$, the conductance is only affected outside of the topological island [see Figs.~\ref{Fig5}(b) and (c)], where the transmission amplitude decreases quadratically, Eq.~(\ref{dotConductance}).

The Andreev processes enabled by the Majorana leakage are a signature of induced superconductivity by the Kitaev chain on the probe-QD. To understand this, we obtain the expression for the DOS in the probe-QD from the Green's Function (GF),

\begin{equation} \label{dotDOS}
    \rho_{dot}(E) = -\frac{1}{\pi}\textrm{Im}\left[ \frac{(E^2 - E_M^2)(E + i \Gamma)}{Z}\right] -\frac{1}{\pi}\textrm{Im}(F),
\end{equation}
where $\rho_{dot} = -\frac{1}{2 \pi} \text{Im}\left[\mathcal{G}_e + \mathcal{G}_h \right]$. The GF of the probe-QD is calculated in App. \ref{App A}, Eqs. (\ref{smatrix}) and(\ref{matrixA}), and has the form 

\begin{equation}
    \mathcal{G}_R^{(dot)} = \begin{pmatrix}\mathcal{G}_e & F \\ F & \mathcal{G}_h \end{pmatrix},
\end{equation}
where $F$ is the anomalous term of the GF and is associated with Andreev processes, as $A_1 = A_{21} = |F|^2 \Gamma^2$. Thus, we identify the first term in Eq. (\ref{dotDOS}) as the contribution from the probe-QD level and the second from the MBS. By taking $E \to 0$, it is possible to understand which energy level mediates the conductance. For instance, considering $E_M \neq 0$, $F = -E t_0^2/\left[E_M^2(\Gamma^2 + \varepsilon_d^2)\right] \to 0$. Consequently, the probability of an incoming electron being reflected or transmitted as a hole vanishes at $E = 0$. Now, if we consider $E_M = 0$, the first term in Eq. (\ref{dotDOS}) goes to zero, while $F \to -i/2\Gamma$, which is followed by the contribution of Andreev processes in the conductance.

\subsection{Finite temperature}

Now we consider finite temperatures, where the energy range relevant to conductance, see Eq. (\ref{I1}), is broadened. Consequently, it is no longer possible to perfectly track the topological islands where $\gamma^2 = 1/2$ as described above. However, we can increase the resolution of the conductance measurement by tuning the probe-QD level, which
allows us to detect the near zero-energy regions depicted in white in Fig. \ref{Fig1}. 

\begin{figure}
	\centering	\includegraphics[width=0.495\textwidth]{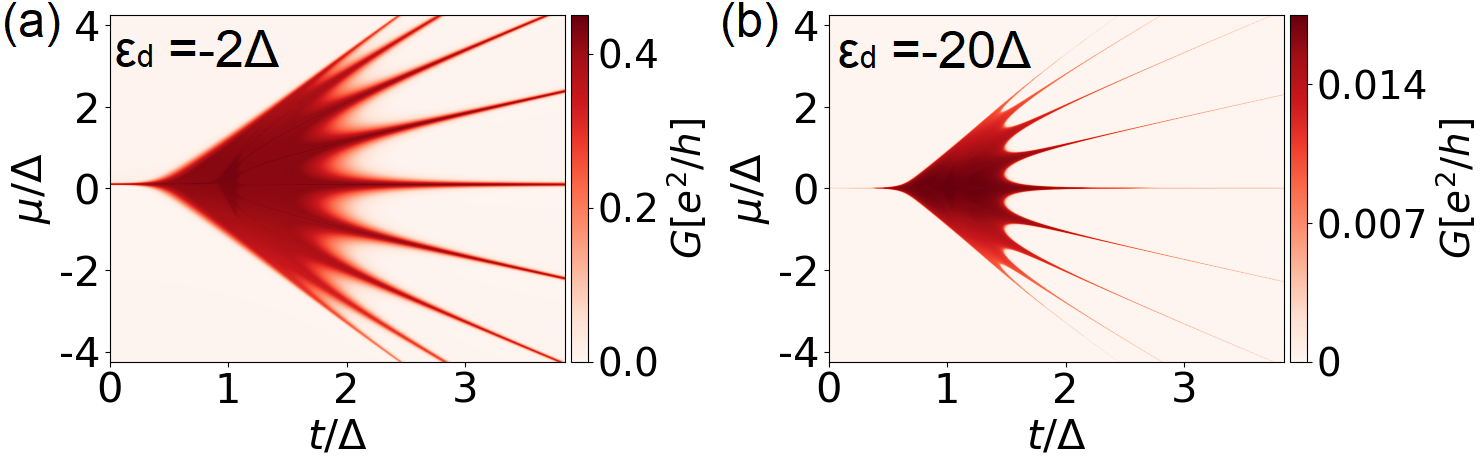}
	\caption{(a) Conductance as a function of $\mu/\Delta$ and $t/\Delta$ for a Kitaev chain with $N = 9$ sites coupled to a probe QD. The zero-bias conductance captures the evolution of the 2-site sweet spot into a region, Fig. \ref{Fig1}(a). Parameters: $\Gamma_1 = \Gamma_2 = 0.1 \Delta$, $T = \Delta/200$, $t_0 = \Delta$, and $\varepsilon_d = -2\Delta$. (b) The precision of the conductance in reproducing the low-energy and high MP regions shown in Fig. \ref{Fig1} improves as we change $\varepsilon_d$ to $-20\Delta$, as the effective coupling of the leads to the Majorana mode is reduced, see Eq. (\ref{effective gamma}).}
	\label{Fig6}
\end{figure}

First, we map the low-energy and high MP region shown in Fig. \ref{Fig1}(d) using the transversal conductance $G$ in Fig. \ref{Fig6}(a), where we vary $\mu/\Delta$ and $t/\Delta$ and show in color the zero-bias conductance for $T = \Delta/200$ and $\alpha = 1/2$. As $E_M$ increases when one moves away from the lines, $G$ rapidly diminishes, reaching its minimum at the middle point between two lines.  In addition, the conductance also captures the number of lines predicted by Eq. (\ref{majoranaLines}) and their convergence at the 2-site sweet spot. 

We observe that by increasing $\varepsilon_d$ [Fig.~\ref{Fig6}(b)] the effective coupling of the MBS with the lead diminishes, improving the resolution when mapping the zero-energy lines given by Eq.~(\ref{majoranaLines}). To explain this result, we redefine the Majorana operator when we couple the Kitaev chain to the probe-QD. For simplicity, we consider $E_M = 0$. We diagonalize the BdG Hamiltonian, Eq.~(\ref{dotKitaevH}), in the subspace of basis $\{c_d , c_d^\dagger , \gamma_1 \}$. We obtain a fermionic excitation of energy $\varepsilon_1 = \sqrt{\varepsilon_d^2 + 2 t_0^2}$ and a zero-mode. The associated operators, obtained using Eq.~(\ref{bogoliubons}), are $b = \xi_1 c_d + \xi_2 c_d^\dagger + \xi_3 \gamma_1 \neq b^\dagger$, and

\begin{equation} \label{New Majorana operator in the dot}
    \bar{\gamma} = \frac{t_0}{\varepsilon_1}\left( c_d + c_d^\dagger\right) - \frac{\varepsilon_d}{\varepsilon_1} \gamma_1 = \bar{\gamma}^\dagger.
\end{equation}
In particular, by setting $\varepsilon_d = 0$, $\bar{\gamma} = \frac{1}{\sqrt{2}} \left(c_d + c_d^\dagger \right)$. The probe-QD has become part of the Majorana wire due to the induced superconductivity. In this case, $c_d$ is analogous to $c_1$ in the Kitaev chain when $t = \Delta$ and $\mu = 0$ \cite{aliceareview}. We now expand $c_d$ in the new basis, so that $c_d = \xi_1^* b + \xi_2 b^\dagger + \frac{t_0}{\varepsilon_1} \bar{\gamma}$. Considering low energies, we neglect the higher energy excitations and approximate $c_d \sim \bar{\gamma}$ \cite{FuTeleportation}. Finally, we obtain the usual tunnel Hamiltonian between a Majorana operator and metallic leads \cite{flensbergMajoranaChain} by rewriting Eq. (\ref{leadsDot}) as

\begin{equation} \label{tunnelingHNewMajoranas}
    H_T =  \sum_{k, i = 1, 2} \Tilde{t}_{k,i} ( d_{k, i}^\dagger - d_{k, i}) \bar{\gamma},
\end{equation}
such that

\begin{equation} \label{effective gamma}
    \Gamma_{eff} = \Gamma \frac{t_0^2}{\varepsilon_d^2 + 2 t_0^2}.
\end{equation}
This result tells us that the effective coupling of the MBS to the lead can be controlled via $\varepsilon_d$. Therefore, for larger values of $\varepsilon_d$, the conductance more sharply follows the Majorana regions, at the expense of diminishing the conductance values for $T > 0$, as shown in Fig. \ref{Fig6}(b). Hence, experiments are limited in tracking the zero-energy plateaus by the precision of the conductance measurement and electron temperature. 

\section{Conclusions} \label{conclusions}
We have investigated how the 2-site Kitaev sweet spot evolves as we increase the number of sites, detailing the crossover between short and long chains. We have shown that the sweet spot becomes a region of low-energy and high MP even for a small number of sites ($N < 10$). We have found robust zero-energy regions for long chains ($N > 20$). We have identified these regions as topological islands due to the protection of the MBSs. These results might motivate future experiments in scaling the number of sites to create, detect, and manipulate MBSs. We have also shown that the zero-energy solutions are associated with self-adjoint operators that follow non-Abelian statistics, $\gamma^2 = 1/2$. In addition, we have tested the robustness of the topological islands against fluctuations in all parameters. We have  demonstrated that the system becomes increasingly unaffected by local Anderson disorder effects as the number of sites increases.  

Finally, we have revisited the well-known setup of a QD connected to the Kitaev chain. We have shown that at $T \to 0$, energy splittings cause the conductance to be mediated solely by the probe-QD level. Conversely, when $E_M=0$ the MBS mediates the conductance, allowing us to associate $G$ to $\gamma^2$. Finally, we have reproduced the white regions of low energy and high MP shown in Fig. \ref{Fig1} via conductance measurements at finite temperatures. By detuning the probe-QD level, we have shown that such a measurement becomes more accurate, although limited by the precision in the conductance measurement and electron temperature.

\section{Acknowledgements}
This work was supported by the São Paulo Research Foundation (FAPESP) Grant No. 2020/00841-9, and from Conselho Nacional de Pesquisas (CNPq), Grant No. 301595/2022-4. R.A.D. would like to thank L. Pupim for fruitful discussions in the early stages of this work.

\appendix
\section{Conductance derivation and Green's function of the QD} \label{App A}

We start the derivation of the s-matrix from Eq. (\ref{smatrixExpression}), where $\mathcal{H}_0$ is the BdG Hamiltonian, which, in the basis $\psi_0 = \begin{pmatrix} c_d & c_d^\dagger & \gamma_1 & \gamma_2 \end{pmatrix}^T$, reads

\begin{equation} \label{H0BdG}
    \mathcal{H}_{0} = \begin{pmatrix}\varepsilon_d & 0 & t_0 & 0 \\ 
    0 & -\varepsilon_d & -t_0 & 0 \\ 
    t_0 & -t_0 & 0 & i E_M \\
    0 & 0 & -i E_M & 0 \end{pmatrix}.
\end{equation}
We obtain matrix $W$ from Eq. (\ref{leadsDot}) in the form $\frac{1}{2}\psi_0^\dagger W \psi_{\textrm{leads}}$, with $\psi_{\textrm{leads}} = \begin{pmatrix} d_1 & d_2 & d_1^\dagger & d_2^\dagger \end{pmatrix}^T$.

\begin{equation} \label{Wmatrix}
    W = \begin{pmatrix}t_1 & t_2 & 0 & 0  \\
    0 & 0 & -t_1 & -t_2 \\
    0 & 0 & 0 & 0 \\
    0 & 0 & 0 & 0 \end{pmatrix}.
\end{equation}
Now we substitute Eqs. (\ref{H0BdG}) and (\ref{Wmatrix}) into Eq. (\ref{smatrixExpression}). We obtain the following expression for the s-matrix, 

\begin{equation} \label{smatrix}
    S = \mathds{1} - i\begin{pmatrix}B_{11} & -B_{12} \\ -B_{21} & B_{22} \end{pmatrix} \otimes \begin{pmatrix}\Gamma_1 & \sqrt{\Gamma_1 \Gamma_2} \\ \sqrt{\Gamma_1 \Gamma_2} & \Gamma_2 \end{pmatrix}.
\end{equation} 

The elements of matrix $B$ are given by

\begin{subequations} \label{matrixA}
\begin{equation}
    B_{11 (22)} = Z^{-1} [E^2 - E_M^2](E + i\Gamma \pm \varepsilon_d) + B_{12},
\end{equation}
\begin{equation}
    B_{12} = B_{21} = - E t_0^2 Z^{-1},
\end{equation}
\begin{equation}
    Z = \left[(E + i\Gamma)^2 - \varepsilon_d^2 \right] [E^2 - E_M^2] - 2 E t_0^2 (E + i \Gamma).
\end{equation}
\end{subequations}
where $\Gamma = (\Gamma_1 + \Gamma_2)/2$. From the above equations, it is immediate to see that if $\Gamma_1 = \Gamma_2$, then $s_{13} = s_{14}$, leading to $A_{1} = A_{21} = |s_{13}|^2$. Similarly, $T_{21} = |s_{12}|^2 = |s_{21}|^2 = T_{12}$.

Now to obtain the conductance, we consider $T_{12} = T_{21}$ and $A_1 = A_{12} = A_{21}$. Therefore, Eq. (\ref{I1}) becomes

\begin{equation}
    I_1 = \frac{e}{h} \int dE \left\{ T_{21} \left[\Tilde{f}(\mu_1) - \Tilde{f}(\mu_2)\right] + A_1 \left[3 \Tilde{f}(\mu_1) + \Tilde{f}(\mu_2)\right] \right\}.
\end{equation}
For small voltage bias, $\Tilde{f} (\mu_i) \sim \mu_i (- \partial f/\partial E)$. Considering also $T \to 0$, so that $-\partial f/\partial E \to \delta(E)$, the current takes the form

\begin{equation} \label{I_L Integration with dfdE}
        I_1 = \frac{e}{h} \int dE \left[ T_{21} (\mu_1 - \mu_2) + A_1 (3 \mu_1 + \mu_2) \right]\left(-\frac{\partial f}{\partial E} \right),
\end{equation}
\begin{equation}
        I_1 = \frac{e^2}{h}V \left[ A_1  (4 \alpha - 1) + T_{21} \right]_{E \to 0} .
\end{equation}
Finally, with $G = dI_1/dV$ we obtain Eq. (\ref{GSimpleForm}).

To evaluate the DOS in the quantum dot, we obtain its retarded GF. To do this, we use the known expression $\mathcal{G}_R = (E - \mathcal{H}_{0} - \Sigma)^{-1}$, where $\Sigma = \Sigma_1 + \Sigma_2$ is the self-energy. The individual self-energy terms are obtained from $\Sigma_i = \sum_k \mathcal{H}_{Ti}^\dagger g_i \mathcal{H}_{Ti}$, where $g_i$ are the GFs of the isolated leads, and $\mathcal{H}_{Ti}$ the matrices coupling the system to the leads, which are given by

\begin{equation}
    \mathcal{H}_{T1(2)} = \begin{pmatrix}t_{1(2)} & 0 & 0 & 0 \\  0 & -t_{1(2)} & 0 & 0 \end{pmatrix}.
\end{equation}
Considering the wide-band limit and symmetric coupling ($\Gamma_1 = \Gamma_2 = \Gamma$), we obtain the following expression for the self-energy

\begin{equation} \label{selfEnergy}
    \Sigma = \begin{pmatrix}-i\Gamma & 0 & 0 & 0 \\
    0 & -i\Gamma & 0 & 0 \\
    0 & 0 & 0 & 0 \\ 0 & 0 & 0 & 0 \end{pmatrix}.
\end{equation}
We point out that for metallic leads $-i\pi W W^\dagger = \Sigma$, which means that $\mathcal{G}_R = (E - \mathcal{H}_0 + i\pi \rho W W^\dagger)^{-1}$, see Eq. (\ref{smatrixExpression}). By substituting Eqs. (\ref{H0BdG}) and (\ref{selfEnergy}) in the GF expression, we obtain

\begin{equation}
    \mathcal{G}_R^{(dot)} = B,
\end{equation}
where the matrix $A$ is described in Eqs. (\ref{matrixA}). The anomalous term of the GF is $F = -t_0^2 E/Z$.

\bibliography{Refs}

\end{document}